\documentclass[sigconf]{acmart}

\usepackage[linesnumbered,ruled,vlined]{algorithm2e}
\usepackage[T1]{fontenc}
\usepackage[export]{adjustbox}
\usepackage{multirow}

\usepackage{booktabs} 

\setcopyright{rightsretained}

\acmDOI{10.1145/nnnnnnn.nnnnnnn}

\acmISBN{978-x-xxxx-xxxx-x/YY/MM}

\acmConference[GECCO '20]{the Genetic and Evolutionary Computation Conference 2020}{July 8--12, 2020}{Cancun, Mexico}
\acmYear{2020}
\copyrightyear{2020}

\acmPrice{15.00}

\usepackage{subfig}
\usepackage{booktabs}
\graphicspath{{./graphics/}}
\DeclareGraphicsExtensions{.pdf,.jpeg,.png}

\usepackage{mathtools}
\DeclarePairedDelimiter\ceil{\lceil}{\rceil}

\begin{document}
\title{Evolutionary Bin Packing for Memory-Efficient Dataflow Inference Acceleration on FPGA}

\author{Mairin Kroes}
\affiliation{\institution{TU Delft}}
\email{m.i.kroes@student.tudelft.nl}

\author{Lucian Petrica}
\affiliation{\institution{Xilinx}}
\email{lucianp@xilinx.com}

\author{Sorin Cotofana}
\affiliation{\institution{TU Delft}}

\author{Michaela Blott}
\affiliation{\institution{Xilinx}}

\renewcommand{\shortauthors}{M. Kroes et al.}

\begin{abstract}
Convolutional neural network (CNN) dataflow inference accelerators implemented in Field Programmable Gate Arrays (FPGAs) have demonstrated increased energy efficiency and lower latency compared to CNN execution on CPUs or GPUs. However, the complex shapes of CNN parameter memories do not typically map well to FPGA on-chip memories (OCM), which results in poor OCM utilization and ultimately limits the size and types of CNNs which can be effectively accelerated on FPGAs. In this work, we present a design methodology that improves the mapping efficiency of CNN parameters to FPGA OCM. We frame the mapping as a bin packing problem and determine that traditional bin packing algorithms are not well suited to solve the problem within FPGA- and CNN-specific constraints. We hybridize genetic algorithms and simulated annealing with traditional bin packing heuristics to create flexible mappers capable of grouping parameter memories such that each group optimally fits FPGA on-chip memories. We evaluate these algorithms on a variety of FPGA inference accelerators. Our hybrid mappers converge to optimal solutions in a matter of seconds for all CNN use-cases, achieve a reduction of up to 65\% in OCM required for deep CNNs, and are up to 200$\times$ faster than current state-of-the-art simulated annealing approaches.
\end{abstract}

%
%

\maketitle


\section{Introduction}

Convolutional Neural Networks (CNNs) have achieved state of the art performance on image classification, object detection, image semantic segmentation and other computer vision tasks and have become an important part of both data-center and embedded workloads. Modern high-accuracy CNNs are typically deep, i.e. they consist of a large number of convolutional layers, each trained through backpropagation. The large number of layers is a key enabler of CNN performance but creates difficulties for their implementation due to the large total number of parameters and the high latency of executing very deep CNNs which makes real-time inference difficult. To reduce inference latency, modern systems typically utilize parallel computing accelerators for CNN inference, either GPUs or Field Programmable Gate Arrays (FPGAs). To date, on FPGA, custom dataflow CNN inference accelerators have achieved the best combination of low latency, high throughput, and low power dissipation \cite{finn_trets}. In the custom dataflow approach, each CNN layer is executed on a dedicated section of the FPGA and its parameters are stored in a dedicated part of the FPGA on-chip memory (OCM), such that the inference process can occur without data ever leaving the FPGA chip, eliminating the high latency and power dissipation associated with external memory reads and writes. 

Of course, a key prerequisite for the custom dataflow approach is for the CNN parameters to fit in FPGA on-chip memory. While quantization and pruning techniques have been successful in reducing the overall size of the CNN parameter memories, one aspect of CNN accelerator design has not been approached in previous work: how to optimally map the diversely-shaped CNN parameter memories to FPGA OCM. In the case of several of the published CNN accelerators, the mapping efficiency is below 70\%, i.e. for structural reasons 30\% of the FPGA OCM bits cannot be utilized. This inefficiency is proportional to inference throughput in frames per second and also increases with the CNN depth.

In this paper we introduce a CNN accelerator memory subsystem construction methodology which enables increased memory mapping efficiency. We achieve this by co-locating multiple CNN parameter memories in a single bank of FPGA OCM, and taking advantage of the multi-port capability of the FPGA memories to minimize the impact to CNN inference performance. Given this design approach, the challenge becomes how to optimally pack CNN parameter memories into available physical memories to achieve the highest memory utilization efficiency within certain throughput constraints. Additionally, given the recent popularity of design space exploration (DSE) techniques for automatically discovering pareto-optimal CNN accelerator configurations~\cite{reggiani2019dsecnnfpga, Motamedi2016DesignSE}, any memory packing algorithm must be very fast to be able to run in the inner loop of a DSE process. Given these considerations, the contributions of this paper are as follows:

\begin{itemize}
    \item We present a novel heuristic which hybridizes Genetic Algorithms and Simulated Annealing with traditional bin packing algorithms to achieve high-efficiency mapping of CNN parameter memories to FPGA OCM
    \item We apply the proposed algorithms on a number of CNN accelerators from previously published work, as well as 3 new accelerators, and demonstrate an increase of up to 65\% in the mapping efficiency, and up to 39\%  reduction in required FPGA OCM size, achieved after running the optimization for under 5 seconds in most cases.
    \item We compare our proposed algorithms against  the state-of-the-art simulated annealing based algorithm in the field and observe 8\% increase in efficiency as well as over 200$\times$ increase in optimization speed.
\end{itemize}

The rest of the paper is structured as follows. Sections \ref{sec:background} and \ref{chap:problem} provide background and state the problem. In section \ref{chap:ga_bin} we present our genetic algorithm and simulated annealing based mapping algorithms. We show in section \ref{sec:Evaluation} that our algorithms improve on previous work in this domain \cite{chow_mempack, ga_bin_pack} in both quality of results (i.e final mapping efficiency) and optimization speed for large dataflow CNN accelerators.

\section{FPGA Accelerated CNN Inference}
\label{sec:background}

\subsection{Accelerator Architectures}
\begin{figure}[]
\centering
\includegraphics[width=0.35\textwidth]{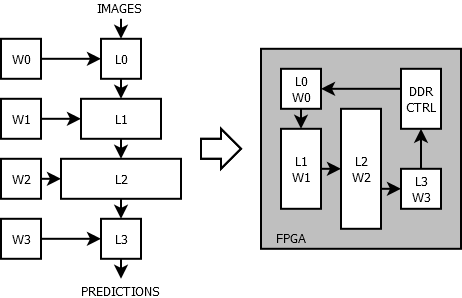}
\caption{CNN Mapped to Dataflow Accelerator on FPGA}
\label{fig:fpga_dataflow}
\vspace{-0.1cm}
\end{figure}

We distinguish two typical approaches of accelerating CNN inference on FPGAs. The first approach leverages a GPU-like matrix of processing engines implemented in FPGA, where the corresponding scheduler determines how to best map the operations of the CNN onto the hardware architecture, typically resulting in a layer by layer compute pattern. The second approach, which is the target of our efforts, leverages feed forward dataflow implementations where the accelerator implements a pipeline of per-layer dedicated compute and associated on-chip parameter memories, as illustrated in Fig. \ref{fig:fpga_dataflow}. All layers of the neural network are in essence spatially unrolled. Benefits are typically lower latency and and higher throughput.
However, as all weights remain in OCM, this becomes the primary bottleneck and limits the layer depth and type of CNN that can be deployed on a single device.

To alleviate this bottleneck, but also to help with the overall computational burden, 
many optimization techniques have been proposed, with quantization and pruning being two of the most popular schemes \cite{han2015deep}.
Quantization is a particularly effective optimization for neural network inference. 
On smaller image classification tasks such as MNIST, SVHN and CIFAR-10, heavily quantized CNNs can achieve significant memory reduction, directly proportional to the reduction in precision, with small loss in accuracy, even when reducing the precision to 1 or very few bits \cite{courbariaux:2016,zhou2016dorefa}.
Furthermore, novel quantization schemes such as~\cite{cai:2017}, and new training and optimization techniques \cite{mishra2017apprentice, zhuang2017s} can potentially recoup the accuracy.
Similarly, pruning can dramatically reduce the size of CNN parameter memory by removing all convolutional filters with \emph{sensitivity} (sum of magnitude of all included weights) below a set threshold.

The progress on quantization and pruning has enabled the implementation of multiple dataflow accelerators and accelerator frameworks \cite{finn_fpga17, finn_trets, ghasemzadeh2018rebnet} for binarized and quantized CNNs in FPGA. Nevertheless, most dataflow accelerators described in previous work still target relatively small binarized CNNs which achieve acceptable accuracy on simple image and speech processing tasks (e.g. classification for MNIST, CIFAR10, SVHN datasets). Dataflow-style FPGA-accelerated binarized CNNs for the Imagenet \cite{deng2009imagenet} 1000-class classification problem have been developed utilizing the FINN \cite{finn_trets} and ReBNet \cite{ghasemzadeh2018rebnet} accelerator frameworks, but have limited Top-1 accuracy compared to equivalent GPU and CPU inference solutions, in the range of 40-50\%. 

To date, achieving state of the art accuracy with dataflow accelerators in FPGA remains a challenge. While approaches such as utilizing higher weight precision, e.g. 2-bit ternary quantization \cite{li2016ternary} instead of binary, or deeper NNs such as ResNet-50 \cite{He2015} have the potential to increase achievable accuracy, they also significantly increase the size of the required on-chip weight storage, making dataflow acceleration difficult.


\subsection{Memory Efficiency vs. Throughput in Dataflow CNNs}
For the remainder of this paper, unless otherwise indicated, we assume FPGA dataflow accelerators are constructed using the architectural principles of the FINN framework~\cite{finn_fpga17}. In FINN-style accelerators, convolutions are lowered to matrix multiplications which are computed by carefully scheduling data to multiply-accumulate (MAC) circuitry on the FPGA fabric. The computational throughput of each layer of the accelerator is controlled by several parallelism variables: the number of vector processing elements (PEs), denoted $N_{PE}$, the vector length of the PE, denoted $N_{SIMD}$, and the number of pixels processed in parallel, denoted $N_{MMV}$. The total number of MAC operations executing in parallel at any given time for any layer is equal the product $N_{PE} \times N_{SIMD} \times N_{MMV}$. To fully utilize the fabric computational resources (Look-up Tables - LUTs - and embedded DSP processors) and therefore maximize throughput, we must perform many MACs in parallel. 

However, this approach forces specific shapes on the parameter memory, in order to achieve parameter readback at the same rate as the compute. Specifically, in each clock cycle and for each layer, the parameter memory must deliver $N_{PE} \times N_{SIMD}$ parameters to the MAC circuits ($N_{MMV}$ is not relevant because pixels share parameters). As such, the memories storing the parameters must have a word width equal to the product $N_{PE} \times N_{SIMD} \times W$, where $W$ is the bitwidth of each parameter. Therefore, as the parallelism (and inference throughput) increase, parameter memories must become wider, and because the total number of parameters for each layer is constant, the depth of the parameter memory must become smaller. In contrast, FPGA OCM consists of block RAM memories (BRAMs) which have a fixed narrow and deep aspect ratio, e.g. 18-bit wide 1024-deep in Xilinx FPGAs. Because of this shape mismatch, CNN parameter memories map inefficiently to BRAM, and given the link between CNN parameter memory shapes and MAC parallelism, high computational throughput implies low BRAM efficiency, and vice-versa.

\begin{figure}
\centering
\includegraphics[width=0.25\textwidth]{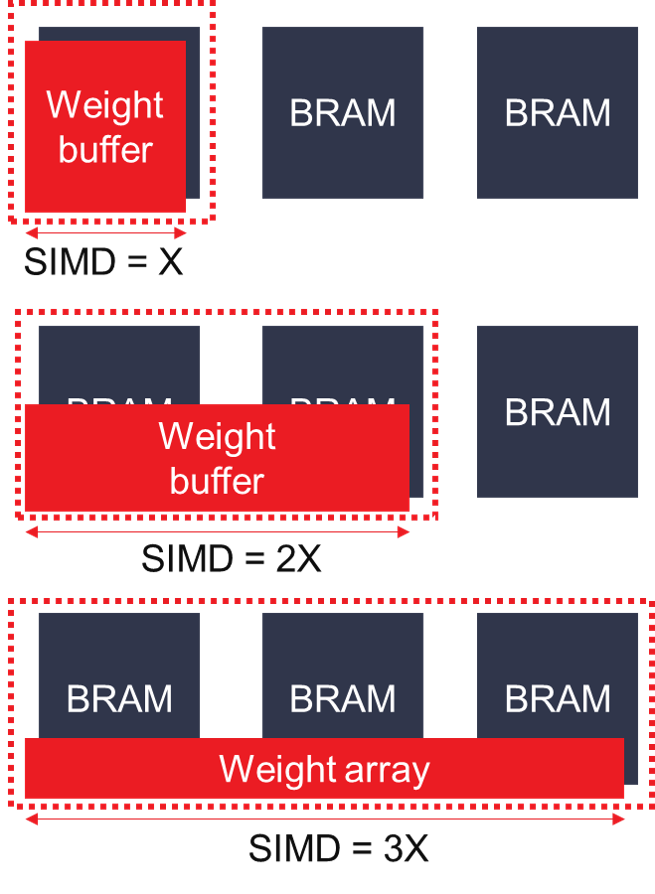}
\caption{Efficiency decreases with increased parallelism}
\label{fig:efficiency_v_simd}
\vspace{-0.2cm}
\end{figure}

Figure \ref{fig:efficiency_v_simd} illustrates this effect. We start from an ideal case of the parameter memory (weight buffer) mapping perfectly to one BRAM. If $N_{SIMD}$ increases by a factor of 2, the shape of the weight buffer must adjust accordingly, and now two adjacent BRAMs must be utilized to each store one half of the buffer. Because the depth has been reduced to half, the efficiency is now 50\%, and can be reduced even further if we increase $N_{SIMD}$ more.

In each case, as the parameter memory becomes wider to provide more parameters in each read cycle, it also becomes shallower and utilizes more BRAMs for implementation. We define the physical RAM mapping efficiency as in Equation \ref{eq:efficiency}, where $D$ is the depth of the parameter memory, $\ceil*{}$ denotes rounding up to nearest integer, and $W_{BRAM}$ and $D_{BRAM}$ denote the width and depth of one BRAM respectively, in bits. Here, the numerator indicates the bits required to be stored and the denominator indicates the total capacity of the BRAM required to actually implement the weight buffer, defined as the product of the width and depth of each physical RAM multiplied by the number of utilized RAMs. The efficiency scales inversely proportional to the exploited parallelism, an undesirable effect.

\begin{equation}\label{eq:efficiency}
E=\frac{N_{PE} \cdot N_{SIMD} \cdot W \cdot D}{W_{BRAM} \cdot D_{BRAM} \cdot \ceil*{\frac{N_{PE} \cdot N_{SIMD} \cdot W}{W_{BRAM}}} \cdot \ceil*{\frac{D}{D_{BRAM}}}}
\end{equation}

Secondly, in the FINN dataflow approach, buffer depth $D$ is proportional to the product of the convolutional kernel size $K$ and the number of channels $C$. $K$ is typically an odd number, most often 3 or 5 in modern CNN topologies,  while $C$ is typically a power of 2 but can be odd if e.g. pruning has been applied to the CNN parameters. Therefore $D$ most of the time does not evenly divide the depth of a physical BRAM, leading to frequent under-fill of the allocated BRAMs.

\subsection{DSE for CNN accelerators}

It is typically the responsibility of a framework-specific resource allocator or design space exploration tool to set the correct value for each parallelism variable in each layer of the CNN dataflow accelerator to maximize overall throughput while remaining within the OCM capacity and LUT/DSP constraints of the target FPGA. Previous work in this area \cite{reggiani2019dsecnnfpga, Motamedi2016DesignSE} has demonstrated that extensive automated search in the design space can identify accelerator configurations better than human designers. As FPGA-accelerated CNNs become deeper and the total number of parallelism variables increases, we expect this trend to continue, as long as appropriate tools exist to quickly estimate the LUT, DSP and OCM requirements of an accelerator from a given set of values of the parallelism variables.

\section{Problem Statement}
\label{chap:problem}
 
Given the strict constraints on the shape of CNN parameter memories, and the fixed shape of FPGA BRAMs, solving the efficiency problem is a matter of finding a way to utilize the space left after one parameter memory has been mapped to a (set of) BRAMs. A straight-forward way to utilize this space is to map a second parameter memory  (or more, if possible) in the empty space. In this approach, the efficiency maximizing problem is analogous to a bin packing problem: the problem of trying to pack a set of objects of varying sizes into bins of certain capacities, with a goal to utilize as few as possible bins to pack all the objects. Here we consider the various CNN parameter memories as the objects, and the physical BRAM instances (or combinations thereof) as the bins into which the objects should be packed. Since this problem is NP-hard, good heuristics are required to obtain acceptable solutions fast. 

The primary factors that make our memory packing problem different from the classical bin packing problem, is that the bins in this case (the block RAM instances) have a limited number of ports which can be used to for parameter reads. For example, if using Xilinx FPGAs, memories have 2 ports, and if we pack 2 parameter memories in one BRAM, we can read one parameter in every clock cycle from each of the packed memories, and the original throughput of the accelerator is maintained. However, beyond 2 parameter memories per BRAM, access to the parameter memories is achieved through time multiplexing of ports, which implies the MAC unit of the accelerators will not be fed parameters in every cycle and the inference throughput suffers. Therefore, beyond simply filling the bin, a good algorithm must also minimize the number of items per bin in order to preserve the inference throughput. In practice, we desire to set an upper limit to the number of items per bin, a so-called cardinality constraint. Secondly, block RAMs can be combined in various ways such that the bins in our case can have variable widths and depths, and therefore have variable capacities. 

These differences significantly deteriorate the efficacy of classical bin packing heuristics that are covered in the literature, since these heuristics build on the concept that an unlimited amount of small items can be used to fill up bins that are almost full. The cardinality constrained version of the bin packing problem was initially explored by Krause et. al. \cite{krause1973} and Kellerer and Pferschy \cite{kellerer1999}. Though despite taking the cardinality constraints explicitly into account, they obtain poor packing results and assume fixed bin sizes, which make these algorithms unsuitable for mapping CNN parameter memories to physical RAMs on FPGA.

Some of the highest performing bin packing algorithms explored in recent literature use genetic algorithms to solve the bin packing problem~\cite{falkenauer1996hybrid, quiroz2015}. In these works, genetic algorithms are combined with classical bin packing heuristics to deliver high quality packing results. More importantly, the proposed strategies utilize an efficient chromosome encoding scheme that was introduced by Falkenauer and Delchambre \cite{ga_bin_pack}. This scheme allows for better exploration of the search space. Since these implementations do not take cardinality constraints into account, some modifications are required before these strategies can be applied to the memory packing problem.

The specific problem of efficient mapping of logical buffers to block RAMs has also been approached in MPack \cite{chow_mempack}, where a simulated annealing algorithm is utilized to discover a good mapping of multiple logical buffers in a single block RAM, but is only demonstrated on relatively small examples compared to modern inference accelerators.
\section{Packing CNN memories to FPGA}
\label{chap:ga_bin}

Previously we established that solving the memory packing problem equates to solving a bin packing problem with a set of hardware constraints, and that genetic algorithms (GA) and simulated annealing (SA) are promising approaches to solve the bin packing problem within the stated constraints. Existing realizations of GA bin packers incorporate recombination techniques that yield good results when there is no upper limit on the amount of items that can be packed into a bin of fixed capacity. Conversely, Vasiljevic and Chow solve the memory packing problem with a simulated annealing approach that explores the search space with random movement of items between bins, referred to as buffer swaps, which can be inefficient for large numbers of bins. We improve on these approaches by introducing \emph{next-fit dynamic}, which is a new heuristic that explicitly takes the custom memory packing related constraints into account and enables a faster and more efficient exploration of the search space. We then embed this heuristic into GA and SA.

\subsection{Next-Fit Dynamic Heuristic}

Next-fit dynamic (NFD) is a recombination technique that is based on the simplest bin packing heuristic next-fit, which has time complexity $O(n)$. As can be seen in Algorithm \ref{alg:first_fit_dynamic}, the NFD heuristic takes as input a list of bins, where each bin contains one or more items. Out of this list we mark the bins that map poorly to BRAM, using an efficiency threshold, decompose the bins into their constituent buffers, and subsequently try to re-pack the buffers into new bins, dynamically adjusting the size of the bin currently being packed according to known BRAM composition rules, e.g. a bin can have widths multiple of $W_{BRAM}$ bits and depths multiple of $D_{BRAM}$.

By design, NFD only adds an additional buffer into an already populated bin if the resulting bin composition leads to less BRAM space being wasted. We allow, however, small admission probabilities ($P_{adm}^w$ and $P_{adm}^h$) that occasionally accept packing configurations that do not immediately improve the mapping efficiencies of the width and height of a bin respectively, to increase the exploration ability of the heuristic and its embedding optimization algorithm.

The NFD strategy enables us to explore large search spaces faster and gives us more control over bin compositions (i.e. not unnecessarily packing buffers if it won't lead to BRAM savings). Moreover, this additional control also enables us to add supplementary restrictions. One example of such a restriction is to exclusively explore bin packing configurations that contain buffers belonging to the same neural network layer (referred to as intra-layer packing), which reduces the average distance between parameter memories and their corresponding MAC circuits on FPGA after the resulting accelerator is implemented, maximizing the operating frequency of the inference accelerator.

\begin{algorithm}
\small
 \KwIn{list of packed bins}
 \KwOut{list of repackaged bins}
 sublist = calculateMapEfficiency(list, threshold);\\
    shuffle(sublist);\\
    \For{buffer in sublist}{
        \eIf{bin height == 0}{
            bin $\gets$ buffer;\\
            update(bin width, bin height);\\
        }
        {
            calculate(new bin height);\\
            gap = calculateGap(BRAM height, bin height);\\ 
            new gap = calculateGap(BRAM height, new bin height);\\
            
            \eIf{length bin < max bin height \textbf{AND}\\ ((new gap < gap \textbf{OR} rnd() < $P_{adm,h}$) \textbf{AND}\\
            (bin width == buffer width \textbf{OR} rnd() < $P_{adm,w}$))}
            {
                bin $\gets$ buffer;\\
                update(bin width, bin height);\\
            }
            {
                list $\gets$ bin;\\
                reset(bin, bin width, bin height);\\
                bin $\gets$ buffer;\\
            }
        }
    }
  \If{length bin > 0}{
  list $\gets$ bin;
  }
 \caption{Next-Fit Dynamic (NFD) Heuristic}
 \label{alg:first_fit_dynamic}
 
\end{algorithm}

\subsection{Genetic Algorithm Bin Packing}

\begin{figure}[]
    \centering
    \includegraphics[trim={6.5cm, 7cm, 4.5cm, 7cm}, clip=true, width=0.45\textwidth]{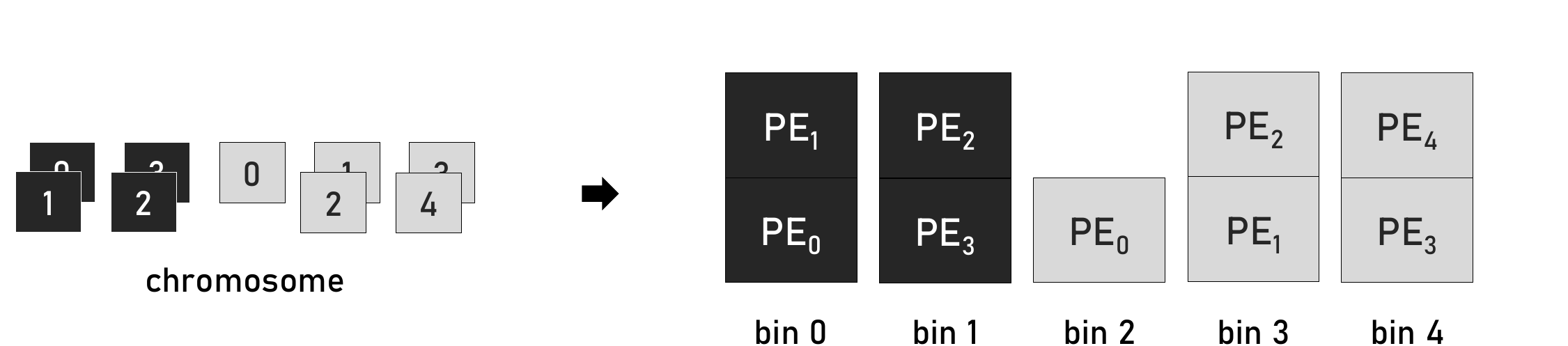}
    \caption{Bin Per Gene Chromosome Encoding}
    \label{fig:bin_gene}
    \vspace{-0.1cm}
\end{figure}

In this work we employ a genetic algorithm that utilizes the so called ``bin per gene'' chromosome representation as illustrated in Figure \ref{fig:bin_gene}. Here a bin refers to a group of CNN parameter memories that will be packed together, so each gene is a list of CNN parameter memories. 



The genetic algorithm pseudocode is listed in Algorithm \ref{alg:ga_pack} and consists of repeating rounds of evolution. In each round, on a given population of bin packing solutions, we apply mutation with a probability $P_{mut}$ for each individual in the population, and then perform fitness evaluation of the population. Using the fitness values, we perform tournament selection where we extract the best solution out of a randomly selected subset of solutions from the current population, and add it to a new population. The selection process is repeated until the new population has the same count as the preceding population, which it replaces and the next evolution round starts.

\LinesNumbered
\begin{algorithm}
\small
 \KwIn{list of partitions, max bin height}
 \KwOut{BRAM cost, list of packed bins}
 initialize(population)\;
 \While{not converged}{
    \For{individual in population}{
        \If{rnd() < $P_{mut}$}{
            mutate(individual);\\
        }
        calculateFitness(individual);\\
    }
    \While{new population count < population count}{
        new individual = tourSelect(population, tour size);\\
        new population $\gets$ new individual
    }
    population = new population;
  }
 \caption{Genetic Algorithm}
 \label{alg:ga_pack}
\end{algorithm}

\paragraph{Mutation}
The mutation operator is the driving factor in the process of exploring the search space. Two different operators are utilized in this work. The first method is the buffer swap method mentioned in \cite{chow_mempack}. Buffer swapping entails moving buffers to different bins, which changes the packing configuration and its associated BRAM cost. The second method is the next-fit dynamic recombination technique - we select a number of genes, unpack the corresponding bins and mark these memories for repackaging with NFD.

\paragraph{Fitness and Selection}

The factor that determines which individual (solution) wins the tournament, is the fitness of that particular individual. In our work we employ a multi-objective fitness function where we compute a weighted sum between BRAM cost and the layer count per bin. Solutions that result in the lowest BRAM cost, and do so with bin configurations that contain buffers from as few as possible different layers are more likely to make it into the next generation. As time progresses, only the solutions that best meet these criteria will remain.

\subsection{Simulated Annealing Bin Packing}
Simulated Annealing is an optimization algorithm first introduced by Kirkpatrick et. al. in \cite{kirkpatrick1983sa}. It is  similar to general hill climbing algorithms, but its distinguishing feature is that it occasionally jumps between hills (i.e. makes large optimization steps) to prevent getting stuck in a local optimum. This escaping behaviour is modeled by random thermal motion that forces the algorithm to sometimes perform (locally) disadvantageous actions. By default, the algorithm accepts an action if it leads to a solution that optimizes a certain cost function. If the action leads to a worse solution, that action might still be accepted with a certain probability $P_A(T)$ as described in Equation \ref{eq:sa_prob}. This probability approaches 1 for high temperatures and decays exponentially as the temperature decreases. As a result, the algorithm will frequently jump between hills at the start of the annealing process, and then selects a hill to climb in the final phase.

\begin{equation}
\begin{aligned}
    P_A(T) \; =\; e^{\frac{-\Delta E}{T}}  
    \label{eq:sa_prob}
\end{aligned}
\end{equation}

Our implementation of the simulated annealing memory packer follows the approach as described in \cite{chow_mempack}, and the general flow is as described in Algorithm \ref{alg:sa_alg}. We first generate a random, yet feasible memory packing solution that adheres to the cardinality constraint. Then we calculate the BRAM cost for this solution. Finally, the optimization process commences as described before. For the different versions of the SA either the simple buffer swap or next-fit dynamic are used to ``perturb'' the solution. If a perturbation was beneficial, the perturbed solution is immediately accepted. Otherwise, the acceptance probability $P_A$ is calculated according to the current temperature, and the acceptance of the bad move might be reconsidered.

\begin{algorithm}
\small
 \KwIn{list of partitions, max bin height, $T_0$, $R_c$}
 \KwOut{BRAM cost, list of packed bins}
 initilize(solution, T);\\
 cost = costFunction(solution);\\
 iter = 0;\\
 \While{not converged}{
    T = calculateTemperature($T_0$,$R_c$,iter);\\
    candidate = perturb(solution);\\
    new cost = costFunction(candidate);\\
    $P_A$ = probability(cost, new cost, T);\\
    \If{new cost < cost \textbf{OR} rnd() < $P_A$}{
        solution = candidate;
        }
    increment iter;\\
  }
 \caption{Simulated Annealing}
 \label{alg:sa_alg}
\end{algorithm} 

In all, we have defined three novel algorithms for solving the CNN parameter memory to FPGA OCM mapping problem: genetic algorithm using buffer swap and NFD as mutation operators, denoted GA-S and GA-NFD respectively, and simulated annealing using NFD as perturbation mechanism, denoted SA-NFD. The simulated annealing with buffer swap, denoted SA-S, has been published in \cite{chow_mempack} but not evaluated for systems of the size of modern CNN inference accelerators.
\section{Evaluation}\label{sec:Evaluation}

\subsection{CNN Use-Cases}

We evaluate our buffer to BRAM mapping algorithms on several CNN-based object detection and classification accelerators selected from previous work and listed in Table \ref{tab:hw_baselines}. The table indicates the source publication for each accelerator and also the shapes and number of parameter memories of each accelerator, which serve as input for our buffer to BRAM packing algorithm.

\paragraph{Small Image Classifiers} CNV-WxAy CNNs belong to the BNN-Pynq\footnote{https://github.com/Xilinx/BNN-PYNQ} suite of object classification accelerators. They are FINN-style \cite{finn_fpga17} FPGA accelerators and target embedded (relatively small) FPGA devices such as the Zynq-7020. CNV-W1A1 utilizes binary (1-bit) quantization \cite{courbariaux2016binarized} while CNV-W2A2 utilizes ternary (2-bit) quantization \cite{li2016ternary}. Both CNNs are trained on the CIFAR-10 \cite{krizhevsky2014cifar} dataset and are able to distinguish between 10 classes of common objects (e.g. birds, cars, dogs, etc.).

\paragraph{Mid-Size Image Classifiers} DoReFaNet and ReBNet are medium-size CNNs trained for object classification on the 1000-class ImageNet \cite{deng2009imagenet} dataset. These CNNs are both quantized versions of AlexNet \cite{krizhevsky2012imagenet}, a popular image classification CNN topology, use binary (1-bit) weights, and consist of 5 convolutional layers and 3 fully-connected layers. However, they differ in the folding factors utilized for their implementation and therefore in the shapes of their weight memories, and as such are treated separately in our evaluation. DoReFaNet was first binarized in \cite{zhou2016dorefa} and implemented in FPGA in \cite{finn_fpga17}. ReBNet was described and implemented in FPGA in \cite{ghasemzadeh2018rebnet} where it is denoted 'Arch3'.

\paragraph{Large Image Classifiers} ResNet-50 \cite{He2015} is a high-accuracy classification CNN designed for high-accuracy image classification on the ImageNet dataset. To our knowledge, no ResNet-50 dataflow implementation currently exists, so we develop a folding solution (i.e. define values for the parallelism variables of each layer) according to the design principles of FINN accelerators \cite{finn_fpga17}, assuming binarized weights and aiming to fit within the LUT capacity of the largest commercially available Xilinx FPGA, the Alveo U250. We also implement larger ResNet variants: ResNet-101 and ResNet-152 which are approximately 2 and 3 times deeper than ResNet-50 respectively but share the overall structure. 

\paragraph{Object Detectors} Tincy-YOLO was first published in \cite{finn_trets} and is a binarized-weight variant of YOLO \cite{redmon2016you}, a popular object detection CNN. It is a fully convolutional design consisting of 9 layers, 6 of which utilize binary weights while two utilize 8-bit weights.

\begin{table*}[t]
\small
\caption{\label{tab:hw_baselines}Baseline dataflow accelerators}
\centering
\begin{tabular}{ccccccc}
\toprule
\textbf{Accelerator:} & \textbf{CNV-W1A1 \cite{finn_fpga17}} & \textbf{CNV-W2A2 \cite{finn_fpga17}} & \textbf{Tincy-YOLO \cite{finn_trets}} & \textbf{DoReFaNet \cite{finn_trets}} & \textbf{ReBNet  \cite{ghasemzadeh2018rebnet}} & \textbf{RN50-W1A2} \\\midrule
\textbf{Memory Shapes}            &    $16\times(32,144,1)$ &    $8\times(16,576,2)$ & $16\times(32,144,1)$ & $136\times(45,72,1)$ & $64\times(54,256,1)$ & $368\times(32,256,1)$ \\
{\footnotesize\textbf{$N_{PE}\times(N_{SIMD},D,W)$}}            &    $16\times(32,288,1)$ &    $8\times(16,1152,2)$   & $25\times(8,320,1)$ & $64\times(34,108,1)$ & $64\times(25,384,1)$ & $32\times(64,256,1)$ \\
            &    $4\times(32,2304,1)$ &    $4\times(1,8192,2)$ & $16\times(32,144,1)$ & $32\times(64,108,1)$ & $64\times(36,384,1)$ & $192\times(64,288,1)$    \\
            &    $4\times(1,8192,1)$ &    $4\times(8,9216,2)$ & $80\times(32,2304,1)$ & $68\times(3,144,1)$ & $64\times(32,576,1)$ & $176\times(32,1024,1)$    \\
            &    $1\times(32,18432,1)$ & $3\times(2,65536,2)$ & & $8\times(8,64000,1)$ & $128\times(64,1152,1)$ & $32\times(64,1024,1)$\\
            & $1\times(4,32768,1)$ & $1\times(8,73728,2)$ & & $4\times(64,65536,1)$ & $40\times(50,2048,1)$ & $96\times(64,1152,1)$          \\
          &$1\times(8,32768,1)$ & & & $8\times(64,73728,1)$ & $128\times(64,2048,1)$ &\\\midrule
\textbf{Total Buffers:}  & 43 & 28 & 137 & 320 & 552 & 896        \\\bottomrule
\end{tabular}
\end{table*}

\subsection{Methodology}

\paragraph{GA Fine-Tuning} We first analyze the effect of population size on the quality of results (packing efficiency) and convergence speed of GA-NFD to pack the ResNet-50, in order to derive guidelines with regard to the optimal population sizes. We evaluate a range of population sizes from 5 to 400, with each experiment repeated 5 times with different random seeds to reduce variability. For each experiment we run the optimization process for 7 minutes, which was empirically determined to ensure convergence for all population sizes under evaluation.

\paragraph{Packing Algorithm Comparison} We compare the GA and SA packing algorithms with and without NFD, in terms of wall-clock time to convergence and quality of results, for each of the accelerators under evaluation. For all algorithms we impose a cardinality constraint of a maximum of 4 parameter memories per physical BRAM. The reported time to convergence is defined as the amount of time it takes each algorithm to attain a packing result that is within 1\% of the discovered minimum. For each convergence experiment, we evaluate 10 different initial random seeds.

\paragraph{Mapping Efficiency Increase} We calculate the efficiency of mapping parameter memories to FPGA OCM for each of the CNN accelerators, targeting a maximum bin height of 4 and utilizing both inter-layer (unconstrained) and intra-layer packing strategies. In this set of experiments, we utilize a single packing algorithm, to be selected from the comparisons described above.

\subsection{Experimental set-up}

We implemented the GA and SA packing algorithms in Python code utilizing the DEAP evolutionary computation library \cite{fortin2012deap} (version 1.3.0). We execute the packing algorithms in single-thread mode on a server equipped with Intel Xeon Silver 4110 CPUs, 128 GB of system RAM, and SSD storage. We measure time using Python's time package.

To enable us to check the BRAM counts of a packing solution in hardware, we implemented in Verilog HDL code a circuit representing a bin, (i.e. a set of assembled BRAMs with associated addressing logic for up to 6 co-located CNN parameter memories) and a Python-based post-processor which takes a packing solution and generates a Vivado 2019.1 project and Block Design consisting of bin instances configured according to the packing solution. We synthesize the resulting Vivado project and compare the post-synthesis BRAM counts to the software-estimated counts. We observe no difference in practice between these measurements.

\section{Results}

\subsection{Effect of GA Population Size}

\begin{figure}[]
\centering
\includegraphics[width=0.4\textwidth]{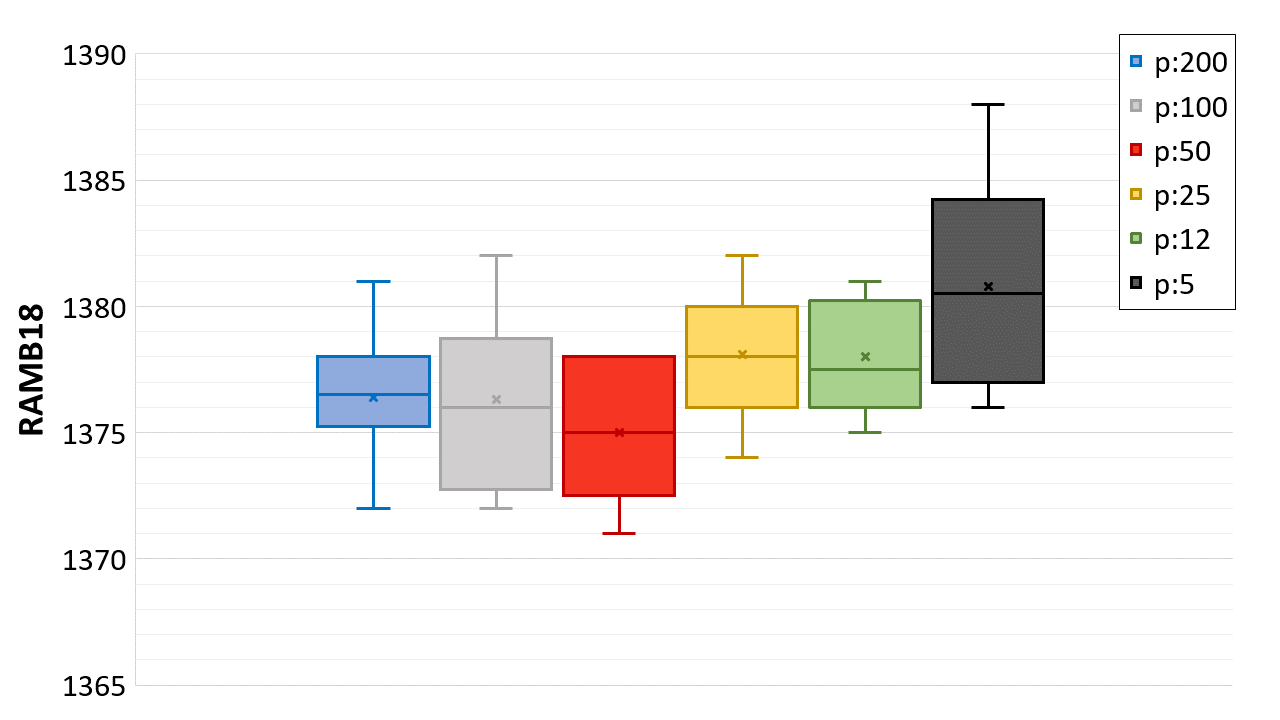}
\caption{QoR comparison for different population sizes on ResNet-50 optimization (GA-NFD)}
\label{fig:generations_boxp}
\vspace{-0.1cm}
\end{figure}

The run-time and QoR (Quality of Results) of genetic algorithms in general depends on the population size utilized. The population size essentially dictates how many candidate solutions are subject to selection and probabilistic mutation at any particular generation.
In Figure \ref{fig:generations_boxp} the results of solving the memory packing problem for ResNet-50 with GA-NFD at varying population sizes are displayed. As can be observed, the algorithm is able to find slightly better results as we scale the population size up to 50. Past this population size we observe a slight regression in performance, however the range of variation in final result after 7 minutes of optimization is very small. Overall, we conclude that population size does not greatly affect the QoR.

\begin{figure}[]
\centering
\includegraphics[width=0.4\textwidth]{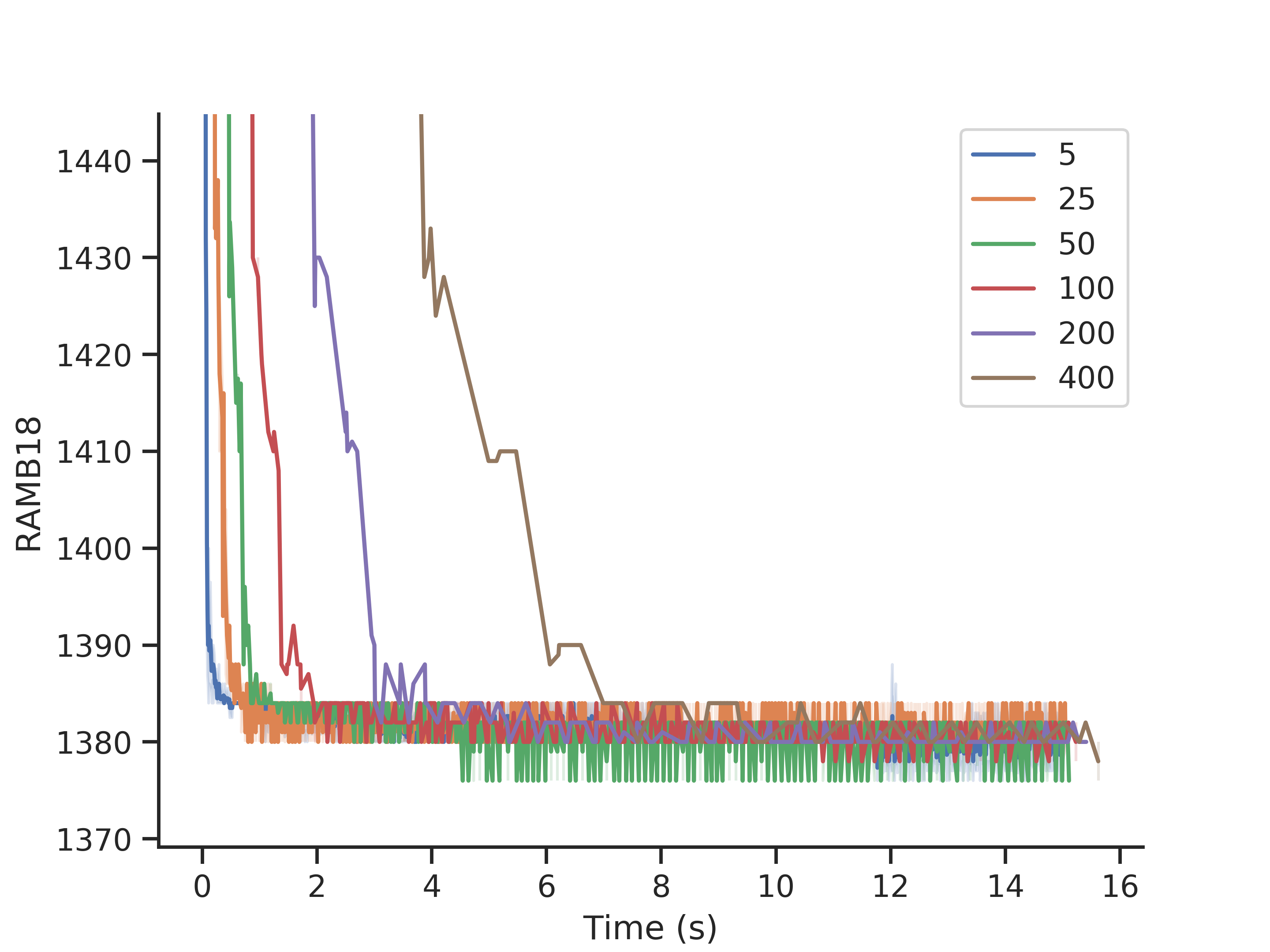}
\caption{Convergence speed for different population sizes on ResNet-50 optimization}
\label{fig:generations_compare}
\vspace{-0.1cm}
\end{figure}

Generally we expect that genetic algorithm experiments utilizing larger population sizes will converge in a smaller number of iterations but those iterations will each be longer in duration than a corresponding iteration for a smaller population size. In this work we are interested in optimizing wall-clock time to solution. To identify the population size that minimizes wall clock time, we pack the respective networks using increasingly larger population sizes and analyze the convergence curves. The results of the population size analysis for ResNet-50 and GA-NFD are illustrated in Fig. \ref{fig:generations_compare}. The best compromise between rapid convergence (in wall-clock time) and quality of results is achieved at a relatively small population size of 50 while the largest population size experiment converged the slowest. This indicates that there is limited benefit from increasing population size to large values. For the experiments performed in this paper, population sizes of approximately 50 appear optimal. 

\subsection{Packing Algorithm Comparison}

In this section the performance of the developed heuristic will be evaluated. As baseline we compare the SA and GA that incorporate next-fit dynamic against SA and GA implementations that use the buffer swap methodology. Both versions of the GA and SA were applied to solve the memory packing problem for the networks as listed in Table \ref{tab:hw_baselines}. We performed extensive hyperparameter tuning for all algorithms to ensure optimal quality of results. The corresponding hyperparameter settings can be found in Table \ref{tab:hyperparams} for simulated annealing and for the genetic algorithm.

\begin{table}[]
\small
\caption{\label{tab:hyperparams}SA and GA Hyperparameters}
\centering
\begin{tabular}{l|ccccc|cc}
\toprule
\multirow{2}{*}{\textbf{Accelerator}} & \multicolumn{5}{c}{\textbf{GA}} & \multicolumn{2}{c}{\textbf{SA}} \\
                                      & $N_p$ & $N_{t}$ & $P_{adm}^{w}$ & $P_{adm}^{h}$ & $P_{mut}$ & $T_0$ & $R_c$\\ \midrule

CNV-W1A1           & 50         & 5 &  0   & 0.1  & 0.3  &  30  & 1             \\
CNV-W2A2           & 50         & 5 &  0   & 0.1  & 0.3  &  30  & 2            \\
Tincy-YOLO         & 75         & 5 &  0   & 0.2  & 0.4  &  30  & 1            \\
DoReFaNet          & 50         & 5 &  0.1 & 0.3  & 0.4  &  30  & 1            \\ 
ReBNet Arch3       & 75         & 5 &  1   & 0.2  & 0.4  &  30  & 1            \\
RN50-W1A2          & 75         & 5 &  0   & 0.1  & 0.4  &  40  & 0.004         \\
RN101-W1A2         & 75         & 5 &  0   & 0.1  & 0.4  &  40  & 0.004         \\
RN152-W1A2         & 75         & 5 &  0   & 0.1  & 0.4  &  40  & 0.004       \\\bottomrule
\end{tabular}
\end{table}

The runtime comparison results can be found in Table \ref{tab:packing_bench} for all networks, with best results highlighted in bold where a clear winner could be distinguished. It has to be mentioned that for the GA the minimum BRAM count of all candidate solutions in a particular generation is tracked. All the algorithm are capable of quickly solving the packing problem for the smaller CNV networks. However, as we increase the problem size (e.g. Tincy-YOLO, DoReFaNet, ResNets) the NFD versions of the algorithms are capable of solving the packing problem much faster, and with higher quality of results. For the ResNets in particular, the NFD algorithms are capable of finding solutions that require up to 8\% less BRAM to implement and reduce the required runtime by a factor of more than 200$\times$ compared to SA-S. GA-S provides poor QoR especially for larger networks and is also slower than all other algorithms. In general, GA-NFD achieves the best QoR while SA-NFD is the faster.

The outlier here is the ReBNet Arch3 accelerator design. This design contains memory partitions with a large variety in widths (SIMD lanes), which causes difficulty for NFD as it is forced to pack together parameter memories with misaligning widths. In order to compete on the metrics as presented in Table \ref{tab:packing_bench} (i.e. BRAM cost and runtime) the hardware constraints had to be relaxed significantly, as is reflected by the high admission probabilities --- $P_{adm}^w$ and $P_{adm}^h$ --- as listed in Table \ref{tab:hyperparams}. Nevertheless, the NFD-based algorithms (especially SA-NFD) arrive at a packing solution significantly faster than buffer swap based GA and SA.

To emphasize the differences in QoR, aside from potentially greater BRAM reductions, the NFD algorithms also provide packing solutions with more ideal bin configurations from a hardware design perspective. The reason for this is that the heuristic typically only packs buffers in bins when it improves the mapping of these bins. As a consequence, the NFD algorithms typically provide packing solutions that contain, on average, bins of lower height, which results in lower throughput penalty.

\begin{table*}
\small
\caption{\label{tab:packing_bench}Memory packing comparison of SA and GA}
\centering
\begin{tabular}{l cc cccc cc cccc}
\toprule
&&& \multicolumn{4}{c}{\textbf{Buffer Swap}} &&& \multicolumn{4}{c}{\textbf{Next-Fit Dynamic}}\\
\cmidrule{4-7} \cmidrule{10-13}
\textbf{Accelerator} &&& $t_{SA-S}$ (s) & $t_{GA-S}$ (s) & $N_{BRAM}^{SA-S}$ & $N_{BRAM}^{GA-S}$ &&& $t_{SA-NFD}$ (s) & $t_{GA-NFD}$ (s) &  $N_{BRAM}^{SA-NFD}$ & $N_{BRAM}^{GA-NFD}$\\ \midrule
CNV-W1A1              &&& 0.1 & 0.2 & 96  & 96 &&& 0.1 & 0.1 & 97  & 96\\
CNV-W2A2              &&& 0.1 & 0.1 & 188 & 190 &&& 0.1 & 0.1 & 190 & 188\\
Tincy-YOLO              &&& 1.8  & 1.7 & 420   & 428 &&& \textbf{0.1}  & 0.2 & 430   & 420\\
DoReFaNet           &&& 1.0  & 1.6 & 3823   & 3826  &&& \textbf{0.1}  & 0.2 & 3849   & \textbf{3794}\\
ReBNet Arch3        &&& 40.1  & 57.5 & \textbf{2301}  & 2313  &&& \textbf{2.2}  & 28.9  & 2483  & 2352\\
RN50-W1A2          &&& 239 & 290 &  1404 & 1472 &&& \textbf{0.8} & 1.7 &  \textbf{1368} & 1374\\
RN101-W1A2           &&& 615 & 935 &  2775 & 3055 &&& \textbf{0.9} & 3.3 & \textbf{2616} & \textbf{2616}\\
RN152-W1A2           &&& 1024 & 1354 & 3864 & 4422 &&& \textbf{1.5} & 4.9 &  3586 & \textbf{3584}\\\bottomrule
\end{tabular}
\end{table*}




\subsection{Achievable Efficiency Increase}

Finally, we applied the memory packing methodology to the accelerators as listed in Table \ref{tab:hw_baselines}, utilizing GA-NFD which achieved the best overall packing performance in Table \ref{tab:packing_bench}. The packing results of the accelerators in original and two different packed configurations are presented in Table \ref{tab:hw_packed}. 

As briefly mentioned before, the term ``intra'' refers to the fact that we only pack buffers corresponding to the same neural network layer together, while in inter-layer packing configurations we do not impose such constraints. The results are presented in terms of BRAM necessary to store the CNN parameters, the resulting mapping efficiency as dictated by Equation \ref{eq:efficiency} and the reduction in memory footprint $\Delta_{BRAM}$. 

\begin{table}
\small
\caption{\label{tab:hw_packed}Mapping Efficiency Increase (GA-NFD)}
\centering
\begin{tabular}{lccc}
\toprule
\textbf{Accelerator} & \textbf{BRAM} & \textbf{Efficiency} & \textbf{$\Delta$\textsubscript{BRAM}}\\ \midrule
CNV-W1A1             &  120             & 69.3\%  &                  \\
CNV-W1A1-Intra       & 100              & 82.3\%  & 1.20$\times$            \\
CNV-W1A1-Inter       & 96               & 86.6\%  & 1.25$\times$  \\ \midrule
CNV-W2A2             & 208              & 79.9\%  &               \\
CNV-W2A2-Intra       & 192              & 86.6\%  & 1.08$\times$    \\
CNV-W2A2-Inter       & 188              & 88.4\%  & 1.11$\times$\\ \midrule
Tincy-YOLO           & 578              & 63.6\%  &             \\
Tincy-YOLO-Intra     & 456              & 80.7\%  & 1.27$\times$ \\
Tincy-YOLO-Inter     & 420              & 87.6\%  & 1.38$\times$  \\\midrule
DoReFaNet            & 4116             & 78.8\%  &             \\
DoReFaNet-Intra      & 3797             & 85.4\%  & 1.08$\times$ \\
DoReFaNet-Inter      & 3794             & 85.5\%  & 1.08$\times$  \\\midrule
ReBNet               & 2880             & 64.1\%  &            \\
ReBNet-Intra         & 2363             & 78.1\%  & 1.22$\times$ \\
ReBNet-Inter         & 2352             & 78.4\%  & 1.22$\times$ \\\midrule
RN50-W1A2            & 2064             & 57.9\%  &         \\
RN50-W1A2-Intra      & 1440             & 82.9\%  & 1.43$\times$  \\ 
RN50-W1A2-Inter      & 1374             & 86.9\%  & 1.50$\times$  \\\midrule
RN101-W1A2            & 4240             & 52.4\% &          \\
RN101-W1A2-Intra      & 2748             & 80.9\% & 1.54$\times$ \\ 
RN101-W1A2-Inter      & 2616             & 84.9\% & 1.62$\times$ \\\midrule
RN152-W1A2            & 5904             & 50.9\% &          \\
RN152-W1A2-Intra      & 3758             & 80.0\% & 1.57$\times$    \\ 
RN152-W1A2-Inter      & 3584             & 83.9\% & 1.65$\times$ \\\bottomrule
\end{tabular}
\end{table}

While the smaller accelerators benefit from the GA-NFD packing, the most benefit is achieved for the ResNet accelerators, which are configured for high throughput and therefore have a low initial memory mapping efficiency roughly around 50\%. We also note that the added constraint of intra-layer mapping does not significantly degrade the achievable efficiency - in most cases the intra-layer efficiency is within 5\% of the inter-layer efficiency.

\section{Discussion and Future Work}

The memory packing methodology presented in this work enables increased memory resource utilization efficiency in modern FPGAs. Our approach is general, i.e. can be utilized for any digital circuit utilizing large parameter memories that are read in predictable fashion at run-time, and is fast compared to the published state-of-the-art. In the specific context of dataflow NN inference accelerators, where memory resource availability is often a design bottleneck, our technique can enable a specific accelerator design to target smaller FPGA devices by becoming more efficient in its OCM usage. The rapid convergence of the NFD-based algorithms to a final packing solution enables their use within CNN accelerator design space exploration frameworks. 

Beyond the CNN acceleration applications presented in this paper, we believe the algorithms presented have a general applicability to FPGA design optimization. Advances can be made by performing multi-objective optimization that takes throughput, memory and LUT utilization of the design into consideration during the evolution process. Thus, evolutionary optimization heuristics can serve to merge many traditional aspects of FPGA electronic design automation which are typically solved in isolation.


\bibliographystyle{ACM-Reference-Format}
\bibliography{references}

\end{document}